\documentclass{elsart}

\usepackage{amsmath,amssymb}
\usepackage{graphicx}
\usepackage{multirow}

\DeclareMathOperator{\diag}{diag}
\DeclareMathOperator{\trace}{Tr}
\DeclareMathOperator{\sign}{sign}

\begin{document}

\begin{frontmatter}

\title{Weak Basis Transformations and Texture Zeros in the Leptonic Sector}

\author[a]{G.~C.~Branco}\ead{gbranco@cftp.ist.utl.pt},
\author[a]{D.~Emmanuel-Costa}\ead{david.costa@ist.utl.pt},
\author[b,a]{R.~Gonz\'{a}lez Felipe}\ead{gonzalez@cftp.ist.utl.pt}\phantom{,}
and
\author[a]{H.~Ser\^odio}\ead{hserodio@cftp.ist.utl.pt}

\address[a]{\small
Departamento de F\'{\i}sica and Centro de F\'{\i}sica Te\'orica de
Part\'{\i}culas (CFTP),
Instituto Superior T\'ecnico, Av. Rovisco Pais, 1049-001 Lisboa, Portugal}

\address[b]{\small
\'Area Cient\'{\i}fica de F\'{\i}sica, Instituto Superior de Engenharia de
Lisboa,\\ Rua Conselheiro Em\'{\i}dio Navarro 1, 1959-007 Lisboa, Portugal}


\begin{abstract}
We investigate the physical meaning of some of the texture zeros which appear in most of the Ans\"atze on leptonic masses and their mixing. It is shown that starting from arbitrary lepton mass matrices and making suitable weak basis transformations one can obtain some of these sets of zeros, which therefore have no physical content. We then analyse four-zero texture Ans\"atze where the charged lepton and neutrino mass matrices have the same structure. The four texture zeros cannot be obtained simultaneously through weak basis transformations, so these Ans\"atze do have physical content. We show that they can be separated into four classes and study the physical implications of each class.
\end{abstract}

\end{frontmatter}

\section{Introduction}

The discovery of neutrino oscillations pointing towards the existence of non-vanishing neutrino masses and large leptonic mixing has rendered the flavour puzzle even more intriguing. There have been many attempts at understanding the pattern of leptonic masses and mixing~\cite{review}, including the introduction of either Abelian or non-Abelian flavour symmetries, some of them leading to texture zeros in the fermion mass matrices. In the leptonic sector there is an extra motivation for introducing texture zeros, namely the fact that without an appeal to theory, it is not possible to fully reconstruct the neutrino mass matrix $m_{\nu}$ from experimental input arising from feasible experiments. It has been shown that this is possible if one postulates the presence of texture zeros in $m_{\nu}$~\cite{Frampton:2002yf} or if one assumes that $\det(m_{\nu})$ vanishes~\cite{Branco:2002ie}.

A difficulty one encounters in an attempt at making a systematic study of experimentally viable texture zeros results from the fact that some sets of these zeros have, by themselves, no physical meaning, since they can be obtained starting from arbitrary fermion mass matrices, by making appropriate weak basis (WB) transformations which leave the gauge currents flavour diagonal~\cite{Branco:1999nb}.

In this paper we investigate in detail what are the texture zeros which can be obtained in the leptonic sector with Majorana neutrinos through WB transformations. We then analyse the physical implications of Ans\"atze where the charged lepton mass matrix $m_{\ell}$ and the effective Majorana neutrino mass matrix $m_{\nu}$ have the same structure (we denote them ``parallel Ans\"atze''), with a total of four independent zeros. These Ans\"atze do have physical meaning, since not all their texture zeros can be simultaneously obtained through WB transformations. Although there is no universal principle requiring parallel structures, they certainly have an easthetical appeal and naturally arise in some classes of family symmetries as well as in the framework of some grand-unified theories~\cite{Ramond:1993kv}.

This letter is organised as follows. In the next section, we show that starting from arbitrary structures for the leptonic mass matrices it is possible to obtain, through WB transformations, $m_{\ell}$ Hermitian with a texture zero in the $(1,1)$ position while $m_{\nu}$ (which is symmetric due to its assumed Majorana nature) has zeros in the $(1,1)$ and $(1,3)$ entries. In section \ref{sec-ZT}, we analyse four-zero parallel Ans\"atze, showing that they can be divided into four different classes. In section \ref{sec-4Z} we confront these Ans\"atze with the present experimental data and analyse their predictions. Finally, in the last section we draw our conclusions.

\section{Creating texture zeros through WB transformations}
\label{sec-ZWB}

We assume the Standard Model with left-handed neutrinos together with some unspecified mechanism leading to lepton-number violation and the generation of a left-handed Majorana mass for neutrinos. The most general WB transformation which leaves the gauge currents invariant is
\begin{equation}
\label{eq:WBT}
\begin{aligned}
m_{\ell}&\longrightarrow m^{\prime}_{\ell}\,=W^{\dagger}\,m_{\ell}\,W_R\,,\\
m_{\nu} &\longrightarrow
m^{\prime}_{\nu}\,=\,W^{\mathsf{T}}\,m_{\nu}\,W\,,
\end{aligned}
\end{equation}
where $W$ and $W_R$ are $3\times3$ unitary matrices, while $m_{\ell}$, $m_{\nu}$ denote the charged lepton and neutrino mass matrices, respectively. It is possible to make a WB transformation which renders $m_{\ell}$ real and diagonal. In this basis, one has:
\begin{equation}
\label{eq:pWB}
\begin{aligned}
m_{\ell}&=D_{\ell}\,,\\
m_{\nu} &=U^{\ast}\,D_{\nu}\,U^{\dagger}\,,
\end{aligned}
\end{equation}
where $D_{\ell}=\diag(m_e,\,m_{\mu},\, m_{\tau})$ and $D_{\nu}=\diag(m_1,\,m_2,\, m_3)$ are real diagonal matrices. The unitary matrix $U$ is the so-called Pontecorvo-Maki-Nakagawa-Sakata (PMNS) matrix~\cite{Pontecorvo:1957cp} which can be parametrised as
\begin{equation}
U=\begin{pmatrix}
c_{12}c_{13} & s_{12}c_{13} & s_{13}e^{-i\delta}\\
-s_{12}c_{23}-c_{12}s_{23}s_{13}e^{i\delta} &
c_{12}c_{23}-s_{12}s_{23}s_{13}e^{i\delta} & s_{23}c_{13}\\
s_{12}s_{23}-c_{12}c_{23}s_13e^{i\delta} &
-c_{12}s_{23}-s_{12}c_{23}s_{13}e^{i\delta}&
c_{23}c_{13}
\end{pmatrix}\,P_{\alpha}\,,
\label{eq:Udef}
\end{equation}
where $c_{ij}\equiv\cos\theta_{ij}$, $s_{ij}\equiv\sin\theta_{ij}$; $\theta_{ij}$ are mixing angles and $\delta$ is CP-violating Dirac phase. The diagonal matrix $P_{\alpha}=\diag(e^{i\alpha_1/2},e^{i\alpha_2/2},1)$ contains the Majorana phases, $\alpha_1$, $\alpha_2$, which have physical meaning only if light neutrinos are Majorana particles. Although Majorana phases do not affect neutrino oscillations, they do play a r\^ole in neutrinoless double beta decay, contributing to so-called effective Majorana mass~\cite{Feruglio:2002af} \begin{equation}
m_{\beta\beta}\,\equiv\,m_1U_{e1}^{\ast\,2}+m_2U_{e2}^{\ast\,2}+
m_3U_{e3}^{\ast\,2}\,.
\end{equation}

\subsection{\bf Creating the $(1,1)$ zero in $m_{\ell}$ and
$m_{\nu}$}

Our goal is to investigate whether it is always possible to find a WB transformation which, starting from arbitrary matrices $m_{\ell}$ and $m_{\nu}$, in the basis given in Eq.~(\ref{eq:pWB}), leads to new matrices $m^{\prime}_{\ell}$ and $m^{\prime}_{\nu}$ such that $(m^{\prime}_{\ell})_{11}=(m^{\prime}_{\nu})_{11}=0$ and $m_{\ell}$ Hermitian. In this case, the WB transformations of Eq.~(\ref{eq:WBT}) are restricted to those with $W_R=W$, \emph{i.e.}
\begin{equation}
\begin{aligned}
m_{\ell}&\longrightarrow m^{\prime}_{\ell}\,=W^{\dagger}\,D_{\ell}\,W\,,\\
m_{\nu} &\longrightarrow
m^{\prime}_{\nu}\,=\,W^{\mathsf{T}}\,U^{\ast}\,D_{\nu}\,U^{\dagger}\,W\,.
\end{aligned}
\end{equation}

The requirement that $(m^{\prime}_{\ell})_{11}$ and $(m^{\prime}_{\nu})_{11}$ vanish leads to the conditions
\begin{align}
m_e\,\left|W_{11}\right|^2+m_{\mu}\,\left|W_{21}\right|^2+m_{\tau}\left|W_{31}
\right|^2=0\,,
\label{eq:za}
\\[2mm]
m_1\,X_{11}^2+m_2\,X_{21}^2+m_3\,X_{31}^2=0\,,
\label{eq:zb}
\end{align}
where $X\equiv U^{\dagger}W$. The matrix elements $X^2_{i1}\, (i=1,2,3)$ in Eq.~(\ref{eq:zb}) are given by
\begin{equation}
\begin{split}
X_{i1}^2 & =  U^{\ast\:2}_{1i}\,W_{11}^2 +
U^{\ast\:2}_{2i}\, W_{21}^2 + U^{\ast\:2}_{3i}\,W_{31}^2 + 2 \: U^{\ast}_{1i}
 W_{11} U^{\ast}_{2i} W_{21} \\  & + 2 \: U^{\ast}_{1i} W_{11} U^{\ast}_{3i}
 W_{31}  + 2 \: U^{\ast}_{2i} W_{21} U^{\ast}_{3i}
 W_{31}\,.
\label{X}
\end{split}
\end{equation}
It is clear that in order for Eq.~(\ref{eq:za}) to have a solution, one of the masses $m_e$, $m_{\mu}$ or $m_{\tau}$ must have a sign opposite to the other two. This requirement can be always fulfilled, since the sign of a Dirac fermion mass can always be changed by making an appropriate chiral transformation. In order for Eq.~(\ref{eq:zb}) to have a solution, the three real non-negative quantities $a_i\equiv|m_i\,X_{i1}^2|$ should be such that a triangle can be formed with sides $a_1$, $a_2$ and $a_3$. A necessary and sufficient condition for them to be the sides of a triangle is that:
\begin{equation}
2\left(
a_1^2\,a_2^2\,+\,a_1^2\,a_3^2\,+\,a_2^2\,a_3^2\right)
\,-a_1^4-a_2^4-a_3^4 \geq 0\,.
\label{eq:tri}
\end{equation}
Given $(m_e,\,m_{\mu},\, m_{\tau})$, $(m_1,\,m_2,\, m_3)$ and $U$, a solution to Eqs.~(\ref{eq:za}) and (\ref{eq:zb}) can be found through the following procedure:
\begin{enumerate}
\item[(i)] Find $|W_{11}|$, $|W_{21}|$ and $|W_{31}|$ such that Eq.~(\ref{eq:za}) is satisfied. It is clear that this is always possible. One can parametrise the first column of $W$ as \begin{equation} |W_{11}|\,=\,\cos\theta\cos\psi\,,\quad|W_{21}|\,=\,\sin\theta\cos\psi\,, \quad  |W_{31}|\,=\,\sin\psi\,.
    \end{equation}
    Then a solution of Eq.~(\ref{eq:za}) can be found by adjusting the angles $\theta$ and $\psi$.
\item[]
\item[(ii)] In order to satisfy Eq.~(\ref{eq:zb}), one has to choose $W$ in such a way that the inequality in Eq.~(\ref{eq:tri}) is verified. Finding a solution of Eq.~(\ref{eq:zb}) is then equivalent to the problem of determining the internal angles of a triangle from the knowledge of its sides. If we denote $\varphi_{ij}\equiv\arg(X_{ij})$, the internal angles of the triangle are given by $2\left(\varphi_{21}-\varphi_{11}\right)$ and $2\left(\varphi_{31}-\varphi_{11}\right)$.
\end{enumerate}

\subsection{\bf Creating an additional zero}

Once the zero in the position (1,1) is obtained, a natural question to ask is whether one can get additional WB zeros while keeping $m_{\ell}$ Hermitian. It can be readily seen that there exists a second WB transformation that keeps $(m^{\prime}_{\ell})_{11}=(m^{\prime}_{\nu})_{11}=0$ and leads either to $(m^{\prime}_{\ell})_{13}=0$ or to $(m^{\prime}_{\nu})_{13}=0$. Such a transformation is defined by the unitary matrix
\begin{equation}
\label{eq:secU}
W=\left(
\begin{array}{ccc}
 1 & 0 & 0 \\ 0 & \cos \theta & -e^{i \varphi}\sin \theta \\ 0 &
e^{-i\varphi}\sin \theta & \cos \theta
\end{array}
\right),
\end{equation}
with $\theta$ and $\varphi$ given by
\begin{equation}
\label{eq:secU1}
\tan \theta = \left|\frac{(m_{\lambda})_{13}}{(m_{\lambda})_{12}}\right|,\quad
\varphi
= \arg (m_{\lambda})_{13}-\arg (m_{\lambda})_{12}\,,\quad \lambda=\ell,\nu\,.
\end{equation}
Similarly, it is also possible to perform a WB transformation analogous to Eq.~(\ref{eq:secU}) such that one obtains the second zero in the position (1,2). In this case, the following relations hold
\begin{equation}
\label{eq:secU3}
\tan \theta = \left|\frac{(m_{\lambda})_{12}}{(m_{\lambda})_{13}}\right|,\quad
\varphi=\arg (m_{\lambda})_{13}-\arg (m_{\lambda})_{12}-\pi\,.
\end{equation}

To illustrate numerically this procedure, we take a tri-bimaximal structure for the PMNS matrix, which as shown by Harrison, Perkins and Scott (HPS)~\cite{Harrison:2002er} is consistent with neutrino oscillation data. In this case, the neutrino mixing matrix has the following form:
\begin{equation}
\label{eq:tbmax}
U_{\text{HPS}}\,=\,\begin{pmatrix}
\sqrt{\frac{2}{3}} & \frac{1}{\sqrt{3}} & 0\\
-\frac{1}{\sqrt{6}} & \frac{1}{\sqrt{3}} &  \frac{1}{\sqrt{2}}\\
-\frac{1}{\sqrt{6}} & \frac{1}{\sqrt{3}} &  -\frac{1}{\sqrt{2}}
\end{pmatrix}\,,
\end{equation}
corresponding to $\nu_3$ bimaximally mixed and $\nu_2$ trimaximally mixed. Since $(U_{\text{HPS}})_{13}=0$, there is no Dirac-type CP violation in this scheme. Note also that there is in addition a $\mu-\tau$ reflection symmetry~\cite{Harrison:2002et}. Making use of the tri-bimaximal mixing matrix $U_{\text{HPS}}$, Eqs.~(\ref{X}) imply the following set of equations:
\begin{equation}
 \begin{array}{c}
\label{eq:sys}
m_e\,|W_{11}|^2+m_{\mu}\,|W_{21}|^2+m_{\tau}\,|W_{31}|^2=0\,, \\[2mm]
|W_{11}|^2\,+\,|W_{21}|^2\,+\,|W_{31}|^2=1\,, \\[2mm]
\frac{1}{3}\,(2\,m_1 + m_2)\,W_{11}^2 \,+\,
\frac{1}{6}\,(m_1 + 2\,m_2)\left(W_{21}+W_{31}\right)^2\hspace*{20mm}\\[1mm]
 \hspace*{10mm} \,+\, \frac{1}{2}\,m_3\,\left(W_{21}-W_{31}\right)^2\,+\,
\frac{2}{3}\,(m_2-m_1)\left(W_{21}+W_{31}\right)\,W_{11}=0\,.
\end{array}
\end{equation}
In order to give a numerical solution to this system of equations, we use the following input values for the lepton masses~\cite{Yao:2006px,Maltoni:2004ei}: \begin{equation}
\begin{aligned}
m_e=0.511\,\text{MeV}\,,\quad m_{\mu}=105.7\,\text{MeV}\,,\quad
m_{\tau}=1.777\,\text{GeV}\,,\\
\Delta m_{21}^2=7.6\times10^{-5}\,\text{eV}^2\,,\quad
\Delta m_{32}^2=2.4\times10^{-3}\,\text{eV}^2\,.
\label{eq:i2}
\end{aligned}
\end{equation}
Choosing $m_{\mu}<0$, $m_1=0.01$ eV, and all other masses positive, we find
\begin{equation}
W_{11}=\,0.869\,,\quad
W_{21}=\,0.480\,e^{1,72\,i}\,,\quad
W_{31}=\,0.116\,.
\end{equation}
If one makes use of the WB transformation defined in Eqs.~(\ref{eq:secU}) and
(\ref{eq:secU1}), the lepton mass matrices become
\begin{equation}
m_{\ell}=\begin{pmatrix}
 0     & 0.093 & 0.19\\
 0.093 & 0.18  & 0.64\\
 0.19  & 0.64  & 1.50
\end{pmatrix}\,\text{(GeV)}\,,\quad
m_{\nu}=\begin{pmatrix}
 0 & 0.023 & 0\\
 0.023 &  0.038  & 0.010\\
   0 &  0.010 &  0.013
\end{pmatrix}\,\text{(eV)}\,.
\label{eq:ex}
\end{equation}
Therefore, Eq.~(\ref{eq:ex}) shows the explicit form of the leptonic mass matrices leading to the HPS structure, in the WB with texture zeros in the elements $(1,1)$ of $m_{\ell}$, and $(1,1)$, $(1,3)$ of $m_{\nu}$.

\section{Four-zero parallel Ans\"atze}
\label{sec-ZT}

In the previous section we have seen that through WB transformations one can obtain three independent texture zeros in the leptonic mass matrices, while maintaining $m_l$ Hermitian. We address now the question whether it is possible to keep $m_{\ell}$ Hermitian and, simultaneously, obtain additional zeros through WB transformations. We will show that this is not possible, thus implying that the assumption of any additional zero does have physical implications.

To be specific, let us investigate whether starting from arbitrary $m_{\ell}$
and $m_{\nu}$ one can make a WB transformation, so that these matrices are put
in the form
\begin{equation}
\label{eq:4tz}
m_{\ell}=\begin{pmatrix}
0 & \ast & 0\\
\ast & \ast & \ast\\
0 & \ast & \ast
\end{pmatrix}\,,
\quad
m_{\nu}=\begin{pmatrix}
0 & \ast & 0\\
\ast & \ast & \ast\\
0 & \ast & \ast
\end{pmatrix}\,,
\end{equation}
with $m_{\ell}$ Hermitian. To prove that this is not possible, let us count the number of independent parameters in these matrices. By making the rephasing
\begin{equation}
m_{\ell}\longrightarrow K^{\dagger}\,m_{\ell}\,K\,,\quad
m_{\nu}\longrightarrow K^{\mathsf{T}}\,m_{\nu}\,K\,,
\end{equation}
with $K\equiv\diag(e^{i\;\varphi_1},e^{i\;\varphi_2},e^{i\;\varphi_3})$, it is straightforward to verify that one can choose $\varphi_i$ so that $m_{\ell}$ becomes real. One has still some freedom left, so that one phase in $m_{\nu}$ is also eliminated. In this way, one is left with four real parameters in $m_{\ell}$ and four real parameters plus three phases in $m_{\nu}$. So altogether one has eleven free parameters in the matrices $m_{\ell}$ and $m_{\nu}$ given in Eq.~(\ref{eq:4tz}). On the other hand, in the leptonic sector with three generations of Majorana neutrinos, there is a total of twelve physical parameters (\emph{i.e.} six lepton masses, three mixing angles and three phases).

From the above simple counting one sees that, in general, the form of Eq.~(\ref{eq:4tz}), with $m_{\ell}$ Hermitian, cannot be obtained through WB transformations, starting from arbitrary $m_{\ell}$ and $m_{\nu}$ matrices. This in turn implies that the Ansatz of Eq.~(\ref{eq:4tz}) does have physical implications. On the other hand, if one relaxes the Hermiticity condition of $m_{\ell}$, one can show that the parallel structure given in Eq.~(\ref{eq:4tz}) can always be obtained through the WB transformation of Eq.~(\ref{eq:WBT}) with $W_R\neq W$. In this case, the total number of free parameters is fourteen and additional assumptions would be necessary to gain predictability\footnote{This is entirely analogous to the case of nearest neighbour interactions (NNI) in the quark sector where the removal of the Hermiticity constraint eliminates any predictability of the NNI Ansatz~\cite{Branco:1988iq}.}. Therefore, in what follows we restrict our analysis to the cases where $m_{\ell}$ is Hermitian. In the quark sector, Ans\"atze analogous to the one of Eq.~(\ref{eq:4tz}) have been considered in the literature~\cite{Branco:1999tw}.

\begin{table}[tb]
\centering
\begin{minipage}{0.6\textwidth}
\centering
\begin{tabular}{cc}
\textbf{Class} & \textbf{Textures} \\
\hline
\multirow{5}{*}{\textbf{I}} &
$\begin{pmatrix}
0 & \ast & 0\\
\ast & \ast & \ast\\
0 & \ast & \ast
\end{pmatrix}
\begin{pmatrix}
0 & 0 & \ast \\
0 & \ast & \ast\\
\ast & \ast & \ast
\end{pmatrix}
\begin{pmatrix}
\ast & 0 & \ast\\
0 & 0 & \ast\\
\ast & \ast & \ast
\end{pmatrix}$
\\ &
$\begin{pmatrix}
\ast & \ast & \ast\\
\ast & 0 & 0\\
\ast & 0 & \ast
\end{pmatrix}
\begin{pmatrix}
\ast & \ast & 0\\
\ast & \ast & \ast\\
0 & \ast & 0
\end{pmatrix}
\begin{pmatrix}
\ast & \ast & \ast\\
\ast & \ast & 0\\
\ast & 0 & 0
\end{pmatrix}$
\\[0.5mm] \hline\textbf{II} &
$\begin{pmatrix}
0 & \ast & \ast\\
\ast & \ast & 0\\
\ast & 0 & \ast
\end{pmatrix}$
$\begin{pmatrix}
\ast & \ast & 0 \\
\ast & 0 & \ast\\
0 & \ast & \ast
\end{pmatrix}$
$\begin{pmatrix}
\ast & 0 & \ast\\
0 & \ast & \ast\\
\ast & \ast & 0
\end{pmatrix}$
\\[0.5mm] \hline \textbf{III} &
$\begin{pmatrix}
0 & \ast & \ast\\
\ast & 0 & \ast\\
\ast & \ast & \ast
\end{pmatrix}
\begin{pmatrix}
0 & \ast & \ast \\
\ast & \ast & \ast\\
\ast & \ast & 0
\end{pmatrix}
\begin{pmatrix}
\ast & \ast & \ast\\
\ast & 0 & \ast\\
\ast & \ast & 0
\end{pmatrix}$
\\[0.2mm] \hline \textbf{IV}\footnote{Class IV is not viable
phenomenologically.} &
$\begin{pmatrix}
\ast & 0 & 0\\
0 & \ast & \ast\\
0 & \ast & \ast
\end{pmatrix}
\begin{pmatrix}
\ast & 0 & \ast \\
0 & \ast & 0\\
\ast & 0 & \ast
\end{pmatrix}
\begin{pmatrix}
\ast & \ast & 0 \\
\ast & \ast & 0\\
0 & 0 & \ast
\end{pmatrix}$
\\[0.5mm] \hline\\[-4mm]
\end{tabular}
\end{minipage}
\caption{All possible four-texture zero Ans\"atze with parallel structure.
Within the same class all the Ans\"atze have the same physical implications.}
\label{tab:ansatz}
\end{table}

\subsection{\bf Weak basis equivalent classes}

It should be noted that different four-zero texture parallel matrices, with zeros located in different positions, may have exactly the same physical content. Indeed, they can be related by a WB transformation, performed by a permutation matrix $P$,
\begin{equation}
\label{eq:P}
 \begin{aligned}
  m^{\prime}_{\ell}&=P^{\mathsf{T}}\,m_{\ell}\,P\,,\\
  m^{\prime}_{\nu}&=P^{\mathsf{T}}\,m_{\nu}\,P\,,
 \end{aligned}
\end{equation}
which automatically preserves the parallel structure, but changes the position of the zeros. The matrix $P$ belongs to the group of six permutations matrices, which are isomorphic to $S_3\,$. The four-zero texture Ans\"atze can then be classified into four classes, as indicated in Table~\ref{tab:ansatz}. Note that such a WB transformation is not allowed in a scheme where one wants to keep the charged lepton matrix diagonal and ordered, as in the Frampton, Glashow and Marfatia (FGM) framework~\cite{Frampton:2002yf}. It is also clear that class IV is not experimentally viable, since it leads to the decoupling of one of the generations.

\subsection{\bf Seesaw realisations of texture zeros}

So far, we have analysed the structure of the effective neutrino mass matrix $m_{\nu}\,$ without considering its origin. It is well known that one of the most attractive scenarios where naturally small neutrino masses are generated is the seesaw framework~\cite{seesaw}. The simplest realisation of this scenario consists of the addition of three right-handed neutrinos to the spectrum of the Standard Model. In this minimal realisation, an effective neutrino mass matrix is generated through the seesaw formula,
\begin{equation}
m_{\nu}=m_D\,M_R^{-1}m^{\mathsf{T}}_D\,,
\end{equation}
where $m_D$ and $m_R$ denote the Dirac neutrino and right-handed Majorana neutrino mass matrices, respectively. This framework is often referred as \mbox{type-I} seesaw mechanism.

In a type-I seesaw, it is reasonable to expect that the presence of family symmetries may lead to texture zeros in $m_D$ and/or $M_R\,$. It is clear that in general these texture zeros do not lead to zeros in $m_{\nu}$. However, there are some special texture-zero structures that, once imposed in $m_D$ and $M_R$, also appear in $m_{\nu}\,$. Among the four classes defined in Table~\ref{tab:ansatz}, only classes I and IV have this remarkable property~\cite{Nishiura:1999yt}. Since class IV is not viable experimentally, it follows that class I is the only viable four-zero texture Ansatz which can be naturally realised in the type-I seesaw framework. On the other hand, classes I, II and III can all be realised in the so-called type-II seesaw, where the dominant contribution to $m_{\nu}$ arises from the coupling of $\nu_{L}$ to a Higgs triplet.

\section{Phenomenological implications of four-zero texture Ans\"atze}
\label{sec-4Z}

We study in this section the phenomenological implications of considering four texture zeros in the leptonic mass matrices with parallel structure, \emph{i.e.} when both matrices $m_{\ell}$ and $m_{\nu}$ have their zeros at the same positions. Since matrices belonging to the same class share exactly the same physical content, we will analyse only three mass matrices $m^{\text{I}}$, $m^{\text{II}}$ and $m^{\text{III}}$ as representatives of the classes~I,~II and III, respectively,
\begin{equation}
m^{\text{I}}=\begin{pmatrix}
0 & \ast & 0\\
\ast & \ast & \ast\\
0 & \ast & \ast
\end{pmatrix}\,,\quad
m^{\text{II}}=\begin{pmatrix}
0 & \ast & \ast\\
\ast & \ast & 0\\
\ast & 0 & \ast
\end{pmatrix}\,,\quad
m^{\text{III}}=\begin{pmatrix}
0 & \ast & \ast\\
\ast & 0 & \ast\\
\ast & \ast & \ast
\end{pmatrix}\,.
\label{eq:3ex}
\end{equation}
We do not consider any example of class IV matrices, since they are
phenomenologically excluded.

In order to have simple analytic relations and most easily address the full complex case, it is convenient to distinguish the phases appearing in the mass matrices as factorisable and non-factorisable. If all the phases from both leptonic mass matrices can be factorised, one can write
\begin{equation}
m_{\ell}\,=\,K_{\ell}\,m^{0}_{\ell}\,K_{\ell}^{\dagger}\,,\quad
m_{\nu}\,=\,K^{\ast}_{\nu}\,m^{0}_{\nu}\,K^{\dagger}_{\nu}\,,
\label{eq:fact}
\end{equation}
where $m^{0}_{\ell}$ and $m^{0}_{\nu}$ are real matrices and $K_{\ell}\,$ and $K_{\nu}$ are diagonal unitary matrices. Note that while for classes I and II the charged lepton mass matrix $m_{\ell}$ is always factorisable in the above sense, this is not true for class III. Indeed, for a charged lepton mass matrix belonging to class III, the quantity $\arg\left[(m_{\ell})_{12} (m_{\ell})^{\ast}_{13}(m_{\ell})_{23}\right]$ is in general non-vanishing. Since this quantity remains invariant under the transformations given in Eq.~(\ref{eq:fact}), the matrix $m^{0}_{\ell}$ would have one remaining phase. When both leptonic matrices are factorisable, the PMNS mixing matrix takes the form
\begin{equation}
U=O^{\mathsf{T}}_{\ell}\, K\, O_\nu\,,
\label{eq:fU}
\end{equation}
where the matrices $O_{\ell}$ and $O_{\nu}$ are real and orthogonal matrices that diagonalise the mass matrices $m^{0}_{\ell}$ and $m^{0}_{\nu}$, respectively. The diagonal unitary matrix $K$, given by $K=K_{\ell}^{\dagger}\,K_{\nu}\,,$ can be parametrised as
\begin{equation}
K=\diag(1, e^{i\phi_1}, e^{i\phi_2})\,.
\label{eq:K12}
\end{equation}

\begin{table}[thb]
\centering\begin{tabular}{l|c|c}
\hline
Parameters & Best fit & $1\sigma$\\
\hline
$\Delta m^2_{21}$ [$10^{-5}$eV${}^2$] & 7.6 & 7.5 -- 7.9\\
$\Delta m^2_{31}$ [$10^{-3}$eV${}^2$] & 2.4 & 2.2 -- 2.5\\
$\sin^2\theta_{12}$ & 0.32  & 0.30 -- 0.34\\
$\sin^2\theta_{23}$ & 0.50  & 0.43 -- 0.57\\
$\sin^2\theta_{13}=|U_{e3}|^2$ & 0.007 & $\leq0.019$\\
\hline
\end{tabular}
\caption{Best-fit values and $1\sigma$ intervals for the three-flavour neutrino
oscillation parameters~\cite{Maltoni:2004ei}.}
\label{tab:expvalues}
\end{table}
To complete our programme and obtain physical predictions, we need to ensure that the PMNS matrix obtained for each class satisfies the experimental constraints presented in Table~\ref{tab:expvalues}, that arise from neutrino oscillation experiments. Since neutrino oscillation experiments probe only the neutrino squared mass differences, $\Delta m^2_{21}$ for solar neutrinos and $\Delta m^2_{31}\simeq\Delta m^2_{32} \gg \Delta m^2_{21}$ for atmospheric neutrinos, we consider in our analysis the two possible hierarchies for the neutrino masses: normal hierarchy, when the mass eigenstate $\nu_3$, separated from $\nu_1$ and $\nu_2$ by the largest mass gap, is the heaviest mass state, and inverted hierarchy, when $\nu_3$ is the lightest mass state.

\subsection{\bf Class I Ansatz}

Let us start with the analysis of Ans\"atze of class I, considering both mass matrices, $m_{\ell}$ and $m_{\nu}$, with all phases factorisable. In this case, they can be parametrised as in Eq.~(\ref{eq:fact}) with
\begin{equation}
m^0_{\lambda}=
\begin{pmatrix}
0 & a_{\lambda} & 0\\
a_{\lambda} & b_{\lambda} & c_{\lambda}\\
0 & c_{\lambda} & d_{\lambda}
\end{pmatrix}\,,\quad \lambda=\ell,\nu\,,
\label{eq:McI}
\end{equation}
where the coefficients $a_{\lambda},b_{\lambda},c_{\lambda}$ and $d_{\lambda}$ are real. Note that the phases in $m_{\ell}$ and $m_{\nu}$ have been absorbed in the matrices $K_{\lambda}$ which will then appear in the PMNS mixing matrix $U$ as indicated in Eq.~(\ref{eq:fU}). Moreover, without loss of generality, the coefficients $a_{\lambda}$, $c_{\lambda}$ and $d_{\lambda}$ can be assumed positive. Taken the coefficient $d_{\lambda}$ as a free parameter, we can express $a_{\lambda}, b_{\lambda}$ and $c_{\lambda}$ in terms of $d_{\lambda}$ and the mass eigenvalues $m_{\lambda\,i}\,(i=1,2,3)$. This is done by using the three weak basis invariants of the matrix $m_{\lambda}$,
\begin{equation}
\begin{aligned}
\trace(m^0)&=m_{1}+m_{2}+m_{3}\,,
\\
\det(m^0)&=m_{1} m_{2} m_{3}\,,
\\
\chi(m^0)&=m_{1}\, m_{2}+m_{1}\,
m_{3}+m_{2}\, m_{3}\,,
\label{eq:invariants}
\end{aligned}
\end{equation}
where the subscript $\lambda$ has been dropped. The coefficients $a$, $b$ and
$c$ read~\cite{Branco:1999nb,Nishiura:1999yt}
\begin{equation}
\begin{array}{l}
a=\sqrt{-\dfrac{m_1 m_2 m_3}{d}}\,,\\
b=m_1+m_2+m_3-d\,,\\
c=\sqrt{-\dfrac{(d-m_1)(d-m_2)(d-m_3)}{d}}\,.
\end{array}
\label{eq:cI}
\end{equation}
From Eqs.~(\ref{eq:cI}) one sees that the range of variation of $d$ depends on the fermion mass signs. Note that the signs of the charged lepton masses have no physical meaning, since they are Dirac-type fermions. However, for Majorana neutrinos, the mass signs have physical meaning since they redefine the Majorana phases in the diagonal matrix $P_{\alpha}$ of Eq.~(\ref{eq:Udef}).

Once the elements of the leptonic matrices are expressed as functions of the leptonic masses and the parameters $d_{\ell}$ and $d_{\nu}$, the real orthogonal matrices $O_{\ell}$ and $O_{\nu}$ are easily constructed for a particular choice of fermion mass signs and a neutrino mass hierarchy. The moduli of the $O_{\ell,\nu}$ elements are given by
\cite{Branco:1999nb,Nishiura:1999yt},
\begin{equation}
|O_{\ell,\nu}| = \left(
\begin{array}{ccc}
 \sqrt{\frac{m_2 m_3 (d-m_1)}{d\, (m_2-m_1) (m_3-m_1)}} &
  \sqrt{\frac{m_1 m_3 (m_2-d)}{d\, (m_2-m_1) (m_3-m_2)}} &
    \sqrt{\frac{m_1 m_2 (d-m_3)}{d\, (m_3-m_1) (m_3-m_2)}} \\
  \\
  \sqrt{\frac{m_1 (m_1-d)}{(m_2-m_1) (m_3-m_1)}} &
   \sqrt{\frac{(d-m_2) m_2}{(m_2-m_1) (m_3-m_2)}} &
   \sqrt{\frac{m_3 (m_3-d)}{(m_3-m_1) (m_3-m_2)}}\\
  \\
  \sqrt{\frac{m_1 (d-m_2) (d-m_3)}{d\, (m_2-m_1) (m_3-m_1)}} &
   \sqrt{\frac{m_2 (d-m_1) (m_3-d)}{d\, (m_2-m_1) (m_3-m_2)}}  &
    \sqrt{\frac{m_3 (d-m_1) (d-m_2)}{d\, (m_3-m_1) (m_3-m_2)}}
\end{array}
\right)\,, \label{matrixO}
\end{equation}
and the corresponding matrix element signs are
\begin{equation}
\sign(O_{\ell,\nu})_{ij} = \left(
\begin{array}{ccc}
 + & + & + \\ \\
\sign(m_1) & \sign(m_2) & \sign(m_3)\\ \\
  \sign(m_2\,m_3) & \sign(m_1) & +\\
\end{array}
\right)\,,
\end{equation}
in the case of normal hierarchy, and
\begin{equation}
\sign(O_{\ell,\nu})_{ij} = \left(
\begin{array}{ccc}
 + & + & + \\ \\
\sign(m_1) & \sign(m_2) & \sign(m_3)\\ \\
\sign(m_3) & + & \sign(m_1\,m_2)\\
\end{array}
\right)\,,
\end{equation}
for an inverted hierarchy. The PMNS matrix $U$ is then obtained using Eq.~(\ref{eq:fU}). To confront the PMNS matrix  with the experimental constraints from neutrino oscillations~\cite{Yao:2006px,Fogli:2005gs,Maltoni:2004ei}, we shall take as input values the charged lepton masses from Eq.~(\ref{eq:i2}) and the neutrino mass squared differences and mixing angles given in Table~\ref{tab:expvalues}.
One of the attractive features of class I Ans\"atze is that the effective Majorana mass $m_{\beta\beta}$ does not necessarily vanish for normal hierarchical neutrinos, although $(m_{\nu})_{11}=0$. Even in the limit where the charged lepton mixing is small, \emph{e.g.} when $d_{\ell}\simeq m_{\tau}$ and $m_e<0$, we have
\begin{equation}
(O_{\ell})_{12}\simeq \sqrt{\frac{|m_e|}{m_{\mu}}}\,,\quad
(O_{\ell})_{13} \simeq (O_{\ell})_{23} \simeq 0\,,
\end{equation}
implying $m_{\beta\beta} \ne 0$, which is in fact a distinctive signature of the four-zero parallel Ans\"atze given in Eq.~(\ref{eq:McI}), contrasting to the Ansatz of case $A_1$ studied by FGM~\cite{Frampton:2002yf}, where $m_{\beta\beta}$ is predicted to vanish.

One can show that within class I Ans\"atze, and in the framework of factorisable phases, an inverted hierarchy for neutrino masses is not consistent with the experimental data, unless the lightest neutrino mass $m_3$ is fine-tuned with the parameter $d_\nu$. This can be understood in two different limits of the PMNS matrix, obtained from Eqs.~(\ref{eq:fU}) and (\ref{matrixO}) with the identification $m_1 \leftrightarrow m_3$. Taking the limit $m_3 \approx 0$, it implies that $|U_{e3}|\approx1$, while in the limit of $|U_{e3}| \approx 0$, and for large $m_3$, we obtain
\begin{equation}
\sin^2\theta_{12}\approx\left|\frac{m_1}{m_1+m_2}\right|\approx\frac12\,,
\end{equation}
unless $m_3 \simeq d_\nu$.

\begin{figure}[t!]
\centering
\begin{tabular}{cc}
\includegraphics[width=0.5\linewidth]{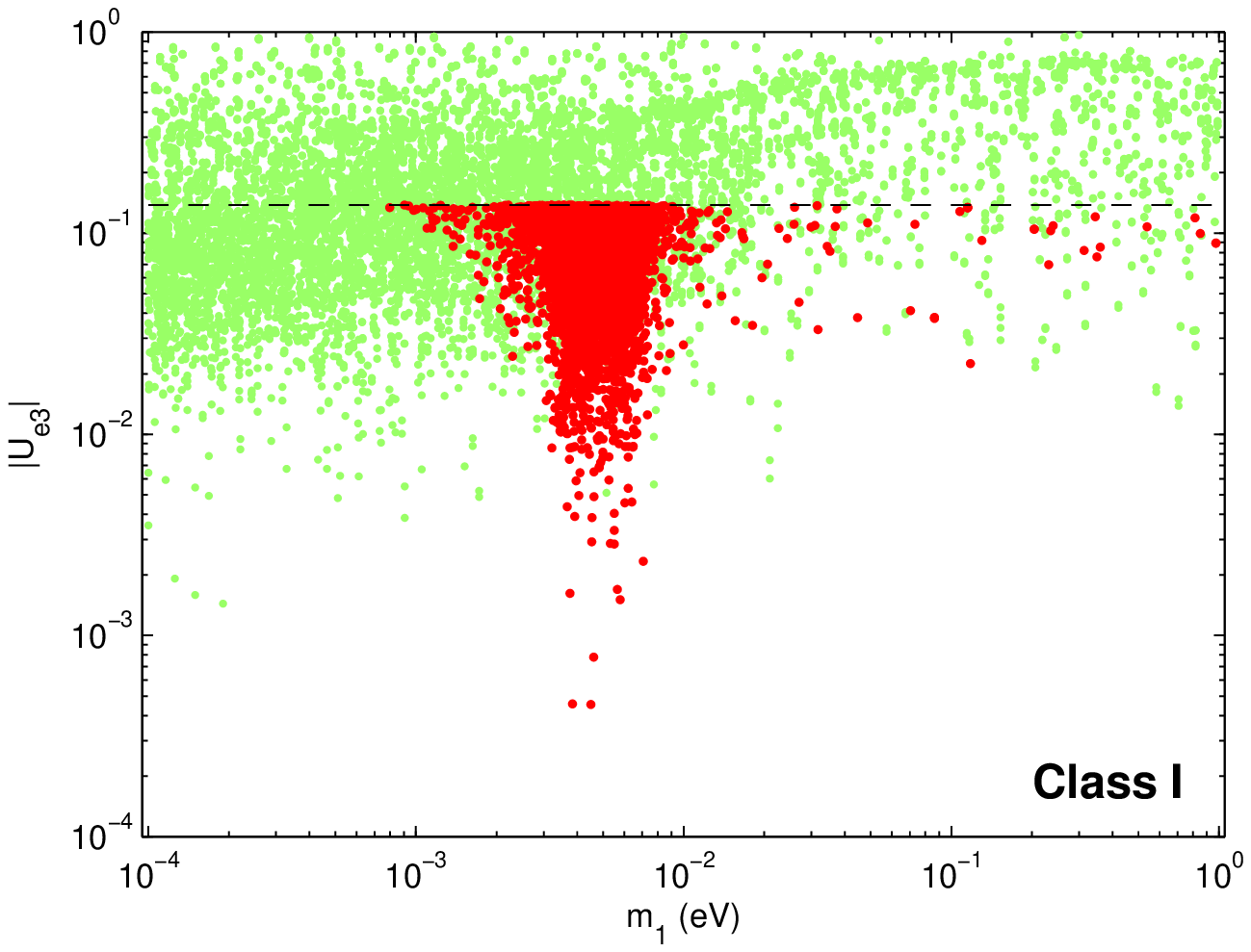} &
\includegraphics[width=0.5\linewidth]{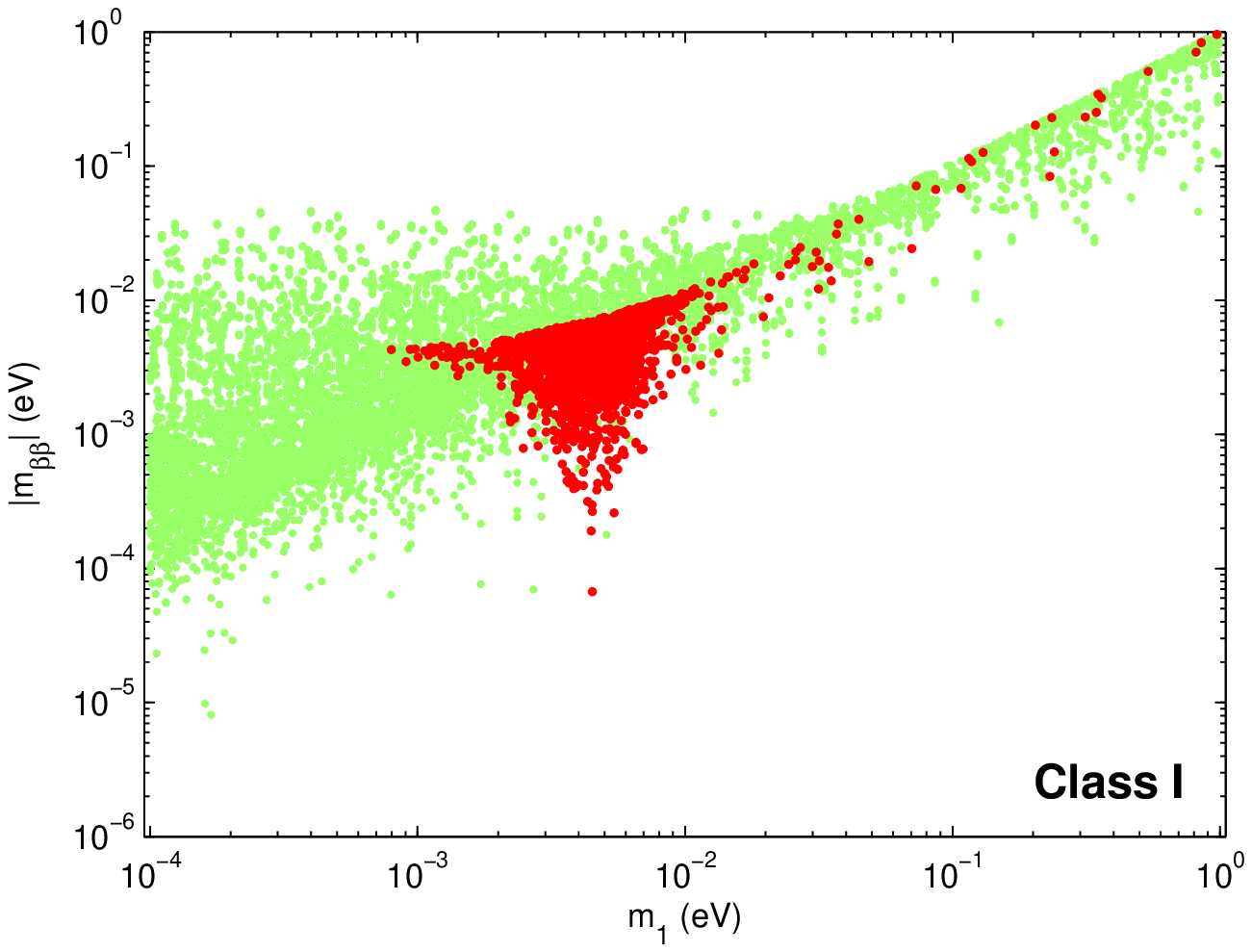}\\
\end{tabular}
\caption{Class I (normal hierarchy): The mixing parameter $|U_{e3}|$ and the
effective
Majorana mass $|m_{\beta\beta}|$ in terms of the lightest neutrino mass $m_1$.
The darker (red) points correspond to points which verify all experimental
constraints. The lighter (green) points define the parameter space allowed by
the Ansatz.}
\label{fig:c1n}
\end{figure}

\begin{figure}[t!]
\centering
\begin{tabular}{cc}
\includegraphics[width=0.5\linewidth]{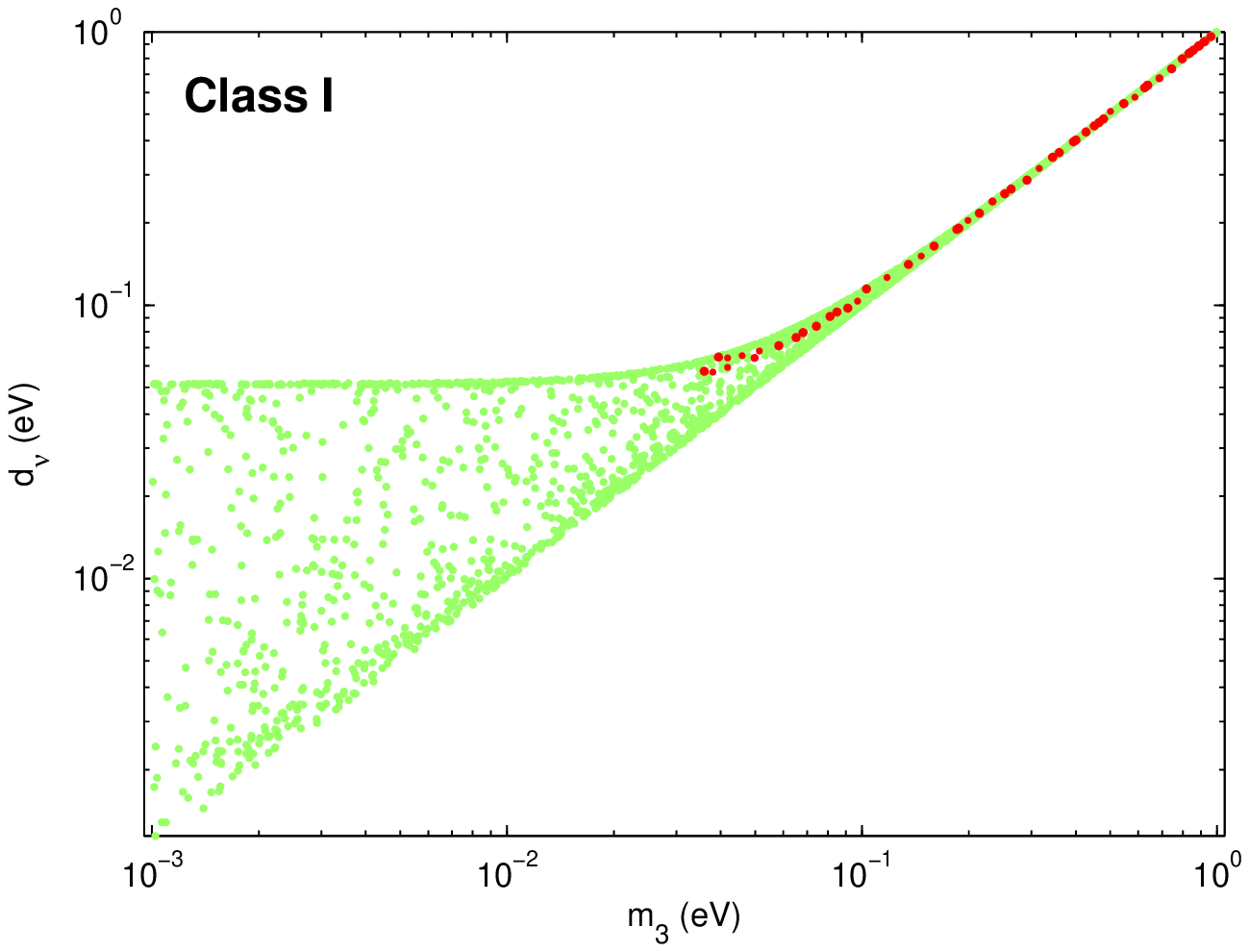} &
\includegraphics[width=0.5\linewidth]{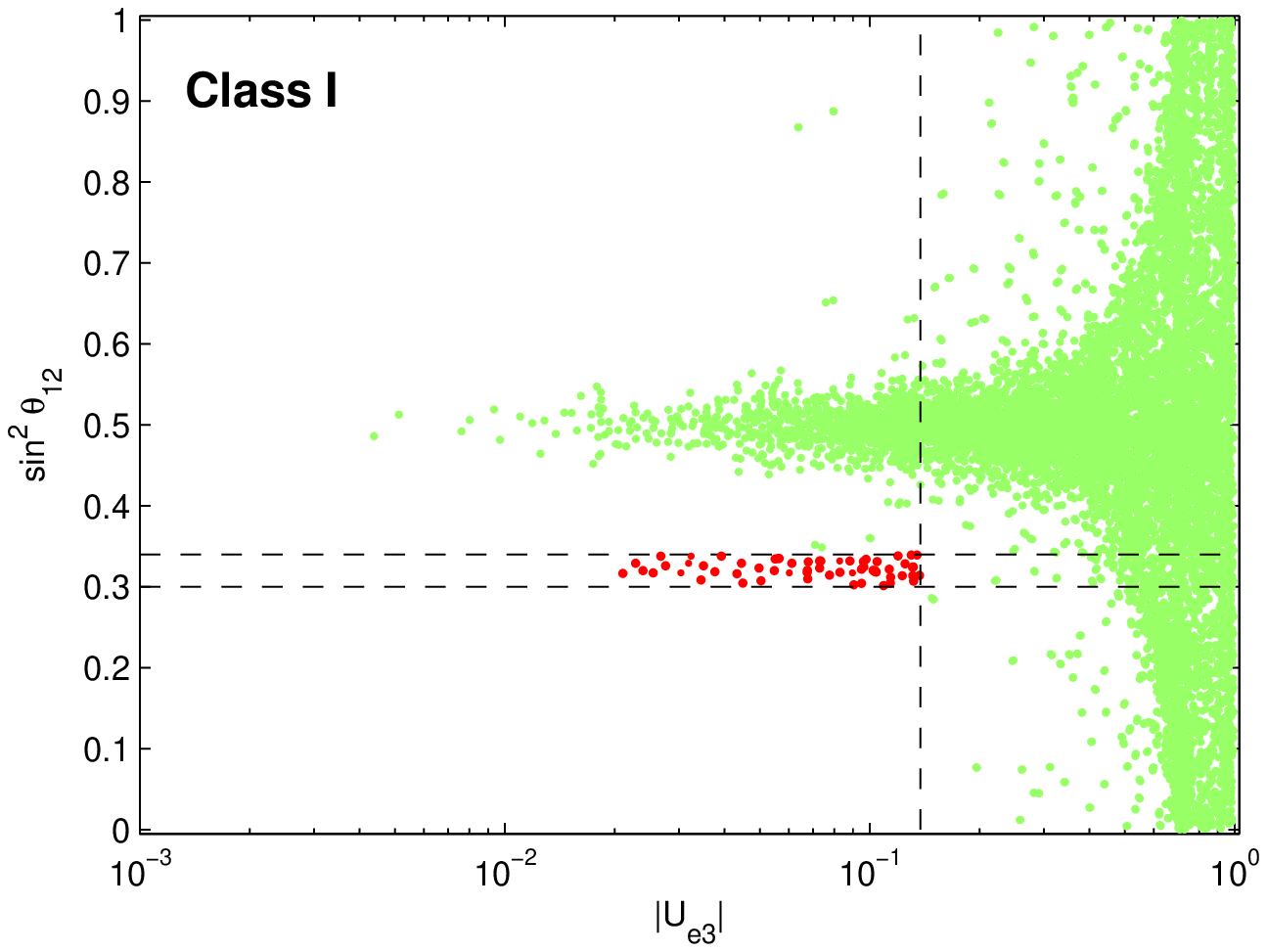}\\
\end{tabular}
\caption{Class I (inverted hierarchy): The Ansatz parameter $d_\nu$ as a function of the lightest neutrino mass $m_3$ and the solar mixing parameter $\sin^2\theta_{12}$ in terms of $|U_{e3}|$. Notice the required fine-tuning ($m_3 \simeq d_\nu$) in order to have solution with inverted hierarchy for this class of Ans\"atze.}
\label{fig:c1i}
\end{figure}

In order to verify whether this result remains valid in the most general case (\emph{i.e.} when not all the phases in $m_{\nu}$ are factorisable), we have proceeded to the numerical analysis of the full complex case, which corresponds, without loss of generality, of having a complex parameter $d_{\nu}$ in the neutrino mass matrix given in Eq.~(\ref{eq:McI}). Notice that three of the four phases in $m_{\nu}$ can be factorized in the sense of Eq.~(\ref{eq:fact}) and their effects absorbed in the PMNS mixing matrix. The results are summarised in Fig.~\ref{fig:c1n} for normal hierarchical neutrinos and in Fig.~\ref{fig:c1i} for an inverted hierarchy. In the latter case, no solution was consistent with the experimental data, except when $d_\nu \simeq m_3 \gtrsim 0.04$, which enforces the analytic procedure.

In the case of a normal hierarchical neutrino mass spectrum, and from the numerical results shown in Fig.~\ref{fig:c1n}, we obtain lower bounds on the lightest neutrino mass $m_1$, the effective Majorana mass $|m_{\beta\beta}|$ and the mixing parameter $|U_{e3}|$, namely, $m_1\gtrsim 8.0\times10^{-4}$~eV, $|m_{\beta\beta}|\gtrsim 7\times10^{-5}\,\text{eV}$ and $|U_{e3}|\gtrsim 4.0\times10^{-4}$. No significant upper bound can be established for any of these quantities; they are only constrained by their present experimental upper limits.

We remark that although the class~I Ansatz given in Eq.~(\ref{eq:McI}) has been previously studied in the context of parallel structures with factorisable phases in both $m_{\ell}$ and $m_{\nu}$ mass matrices~\cite{Nishiura:1999yt,Xing:2003zd}, a complete analysis has not been done. We have presented here a full complex analysis, including all the physical phases.

\subsection{\bf Class II Ansatz}

To analise this class we follow the same procedure as for class I Ans\"atze. When all phases are factorisable, the charged lepton and neutrino mass matrices  can be parametrised according to Eq.~(\ref{eq:fact}), where
\begin{equation}
m^0=
\begin{pmatrix}
0 & a & c\\
a & b & 0\\
c & 0 & d
\end{pmatrix}
\label{eq:McII}
\end{equation}
is a real matrix, and $a$, $c$ and $d$ are positive. Using Eqs.~(\ref{eq:invariants}), we obtain the following relations:
\begin{equation}
\begin{aligned}
a&=\sqrt{-\frac{(m_1+m_2-d) (m_1+m_3-d)
   (m_2+m_3-d)}{m_1+m_2+m_3-2d}}\,,\\[2mm]
b&=m_1+m_2+m_3-d\,,\\[2mm]
c&=\sqrt{\frac{(d-m_1)(d-m_2)(d-m_3)}{m_1+m_2+m_3-2d}}\,.
\end{aligned}
\label{eq:cII}
\end{equation}
The set of relations given in Eq.~(\ref{eq:cII}) allows the construction of the real and orthogonal matrices $O_{\ell}$ and $O_{\nu}\,$, so that the PMNS matrix $U$ can be determined as a function of the six lepton masses, the two parameters $d_{\ell}$, $d_{\nu}$ and the two phases $\phi_1$, $\phi_2$ defined in Eq.~(\ref{eq:K12}).

\begin{figure}[t]
\centering
\begin{tabular}{cc}
\includegraphics[width=0.5\linewidth]{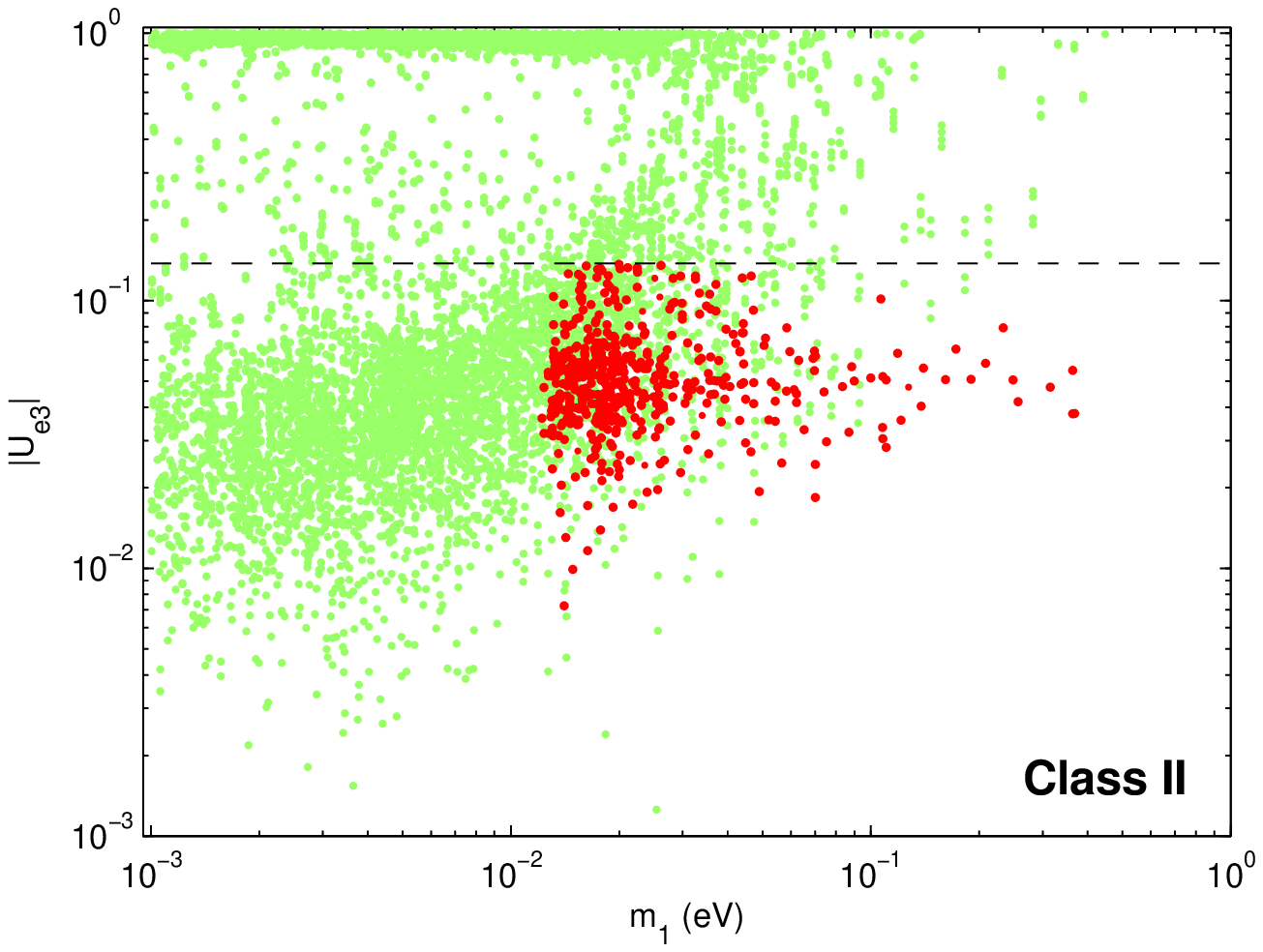} &
\includegraphics[width=0.5\linewidth]{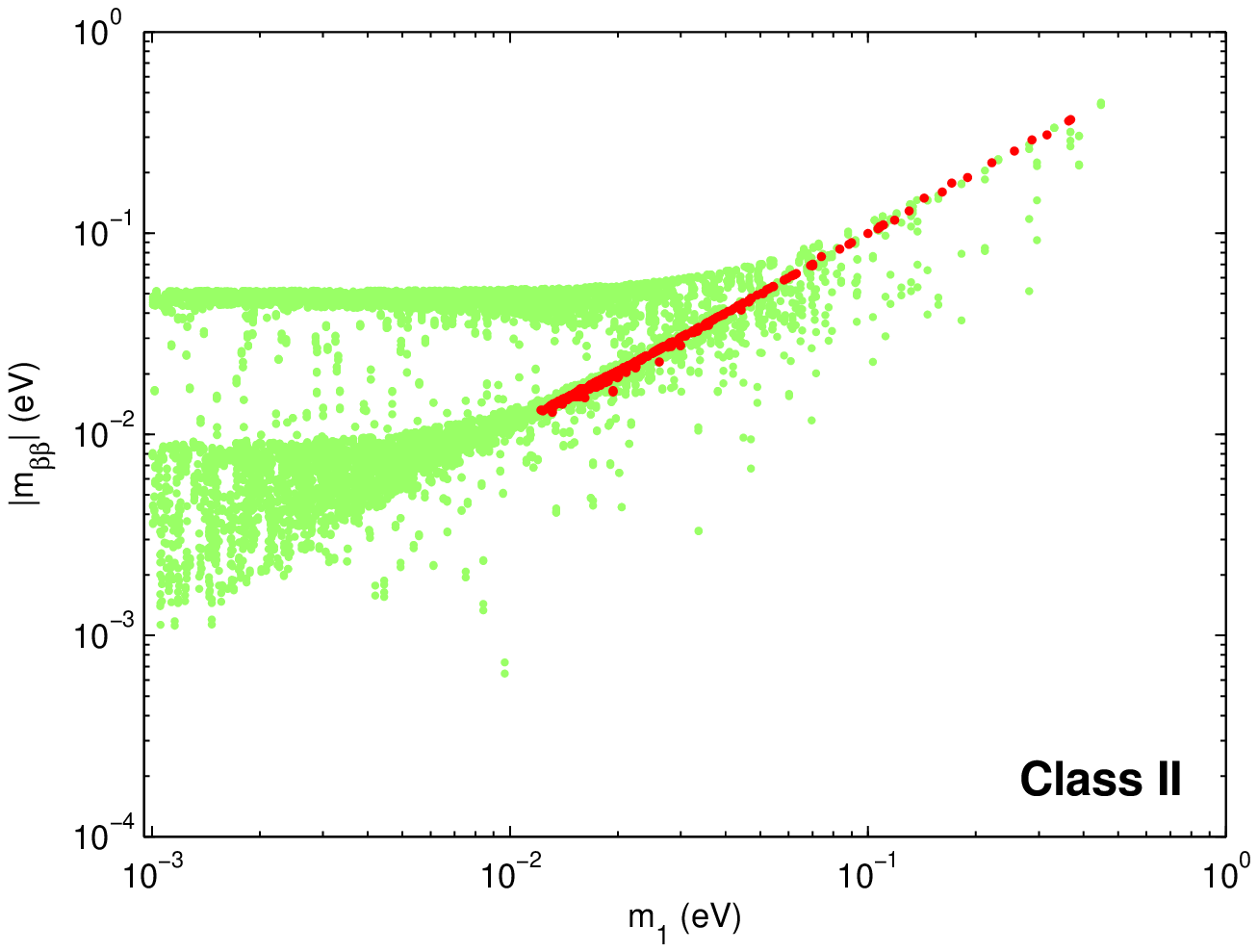}\\
\end{tabular}
\caption{Class II (normal hierarchy): The quantities $|U_{e3}|$ and $|m_{\beta\beta}|$ as functions of the lightest mass $m_1$. The darker (red) points verify all the experimental constraints.}
\label{fig:c2n}
\end{figure}

\begin{figure}[t]
\centering
\begin{tabular}{cc}
\includegraphics[width=0.5\linewidth]{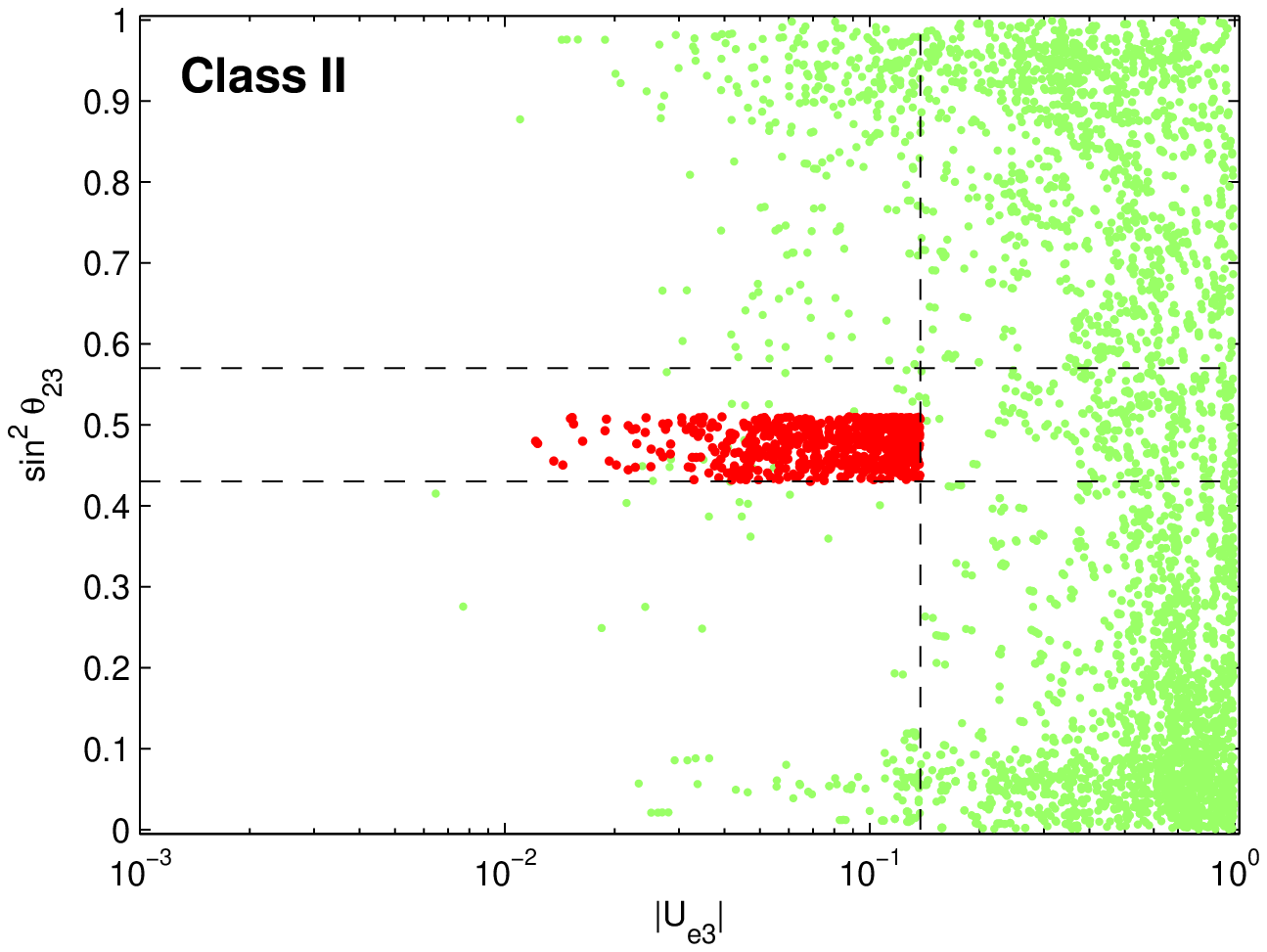} &
\includegraphics[width=0.5\linewidth]{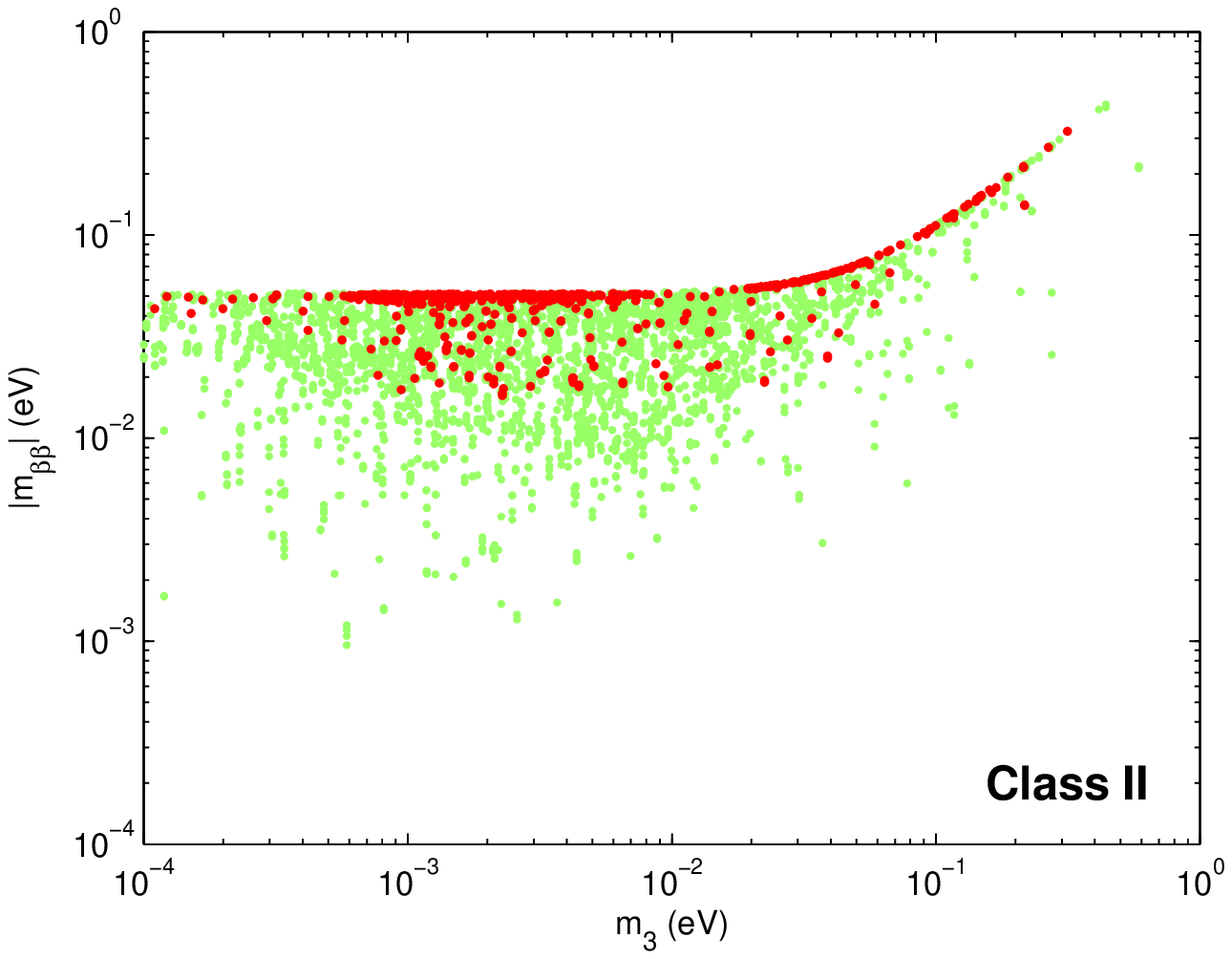}\\
\end{tabular}
\caption{Class II (inverted hierarchy): The atmospheric mixing parameter $\sin^2\theta_{23}$ as a function of $|U_{e3}|$ and $|m_{\beta\beta}|$ in terms of the lightest neutrino mass $m_3$. The darker (red) points verify all the experimental constraints.}
\label{fig:c2i}
\end{figure}

The numerical results for Class II Ans\"atze were obtained within the full complex case, including all physical phases. We can see from Figs.~\ref{fig:c2n} and~\ref{fig:c2i} that both neutrino spectra are allowed by the experimental data. The lightest neutrino mass $m_1$ in the case of normal hierarchy is bounded from below, $m_1 \gtrsim 10^{-2}\,\text{eV}$, while for the inverted hierarchy no bounds can be established. In both cases, the lightest mass is constrained from above only by its present experimental upper limit.

The quantity $m_{\beta\beta}$ is different from zero in both types of neutrino hierarchies. In the case of normal hierarchy we obtain the lower bound $|m_{\beta\beta}| \gtrsim 10^{-2}\,\text{eV}$, while for an inverted hierarchy one has $|m_{\beta\beta}|\gtrsim 2\times10^{-2}\,\text{eV}$. Such values of $m_{\beta\beta}$ could in principle be tested in future neutrinoless double beta decay experiments~\cite{KlapdorKleingrothaus:2004wj,Ejiri:1999rk}. The mixing parameter $|U_{e3}|$ is also bounded from below: $|U_{e3}|\gtrsim 7 \times10^{-3}$ for a normal hierarchy and $|U_{e3}| \gtrsim 10^{-2}$ for an inverted hierarchy.

\subsection{\bf Class III Ansatz}

In the case of factorisable phases, the leptonic mass matrices belonging to this class can be written as in Eq.~(\ref{eq:fact}), where the matrices $m^0_{\lambda}$ are parametrised by real parameters $a$, $b$, $c$ and $d$ in the form
\begin{equation}
m^0=\begin{pmatrix}
0 & a & b\\
a & 0 & c\\
b & c & d
\end{pmatrix}\,.
\label{eq:McIII}
\end{equation}
Here we choose $a$ to be the free parameter, since $d$ is already fixed by the trace invariant in Eq.~(\ref{eq:invariants}),
\begin{equation}
d=m_1+m_2+m_3\,.
\label{eq:c3T}
\end{equation}
The remaining two parameters, $b$ and $c$, are given as functions  of the parameter $a$ and the lepton masses $m_i$,
\begin{equation}
(b\pm c)^2=-(m_{1}\, m_{2}+m_{1}\,
m_{3}+m_{2}\, m_{3})-a^2\,\pm\,\frac{a^2d+m_1m_2m_3}{a}\,.
\label{eq:cIII}
\end{equation}
The set of relations given in Eqs.~(\ref{eq:c3T}) and (\ref{eq:cIII}) allows the construction of the real and orthogonal matrices $O_{\ell}$ and $O_{\nu}$, which in turn determine the PMNS matrix $U$. In this case, $U$ is a function of the six lepton masses, the parameters $a_{\ell}$, $a_{\nu}$ and the two phases $\phi_1$, $\phi_2$ defined in Eq.~(\ref{eq:K12}).

\begin{figure}[t]
\centering
\begin{tabular}{cc}
\includegraphics[width=0.5\linewidth]{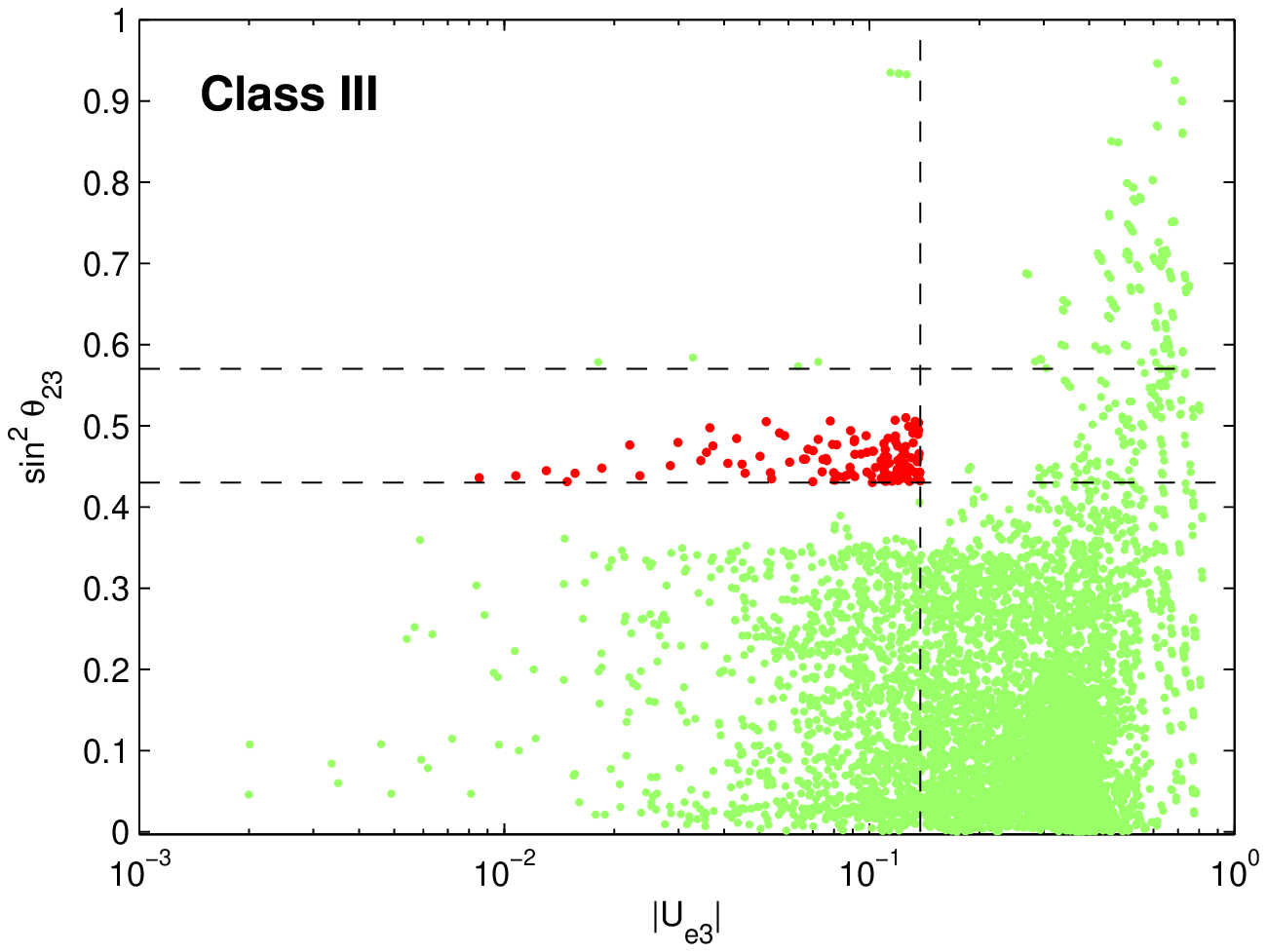} &
\includegraphics[width=0.5\linewidth]{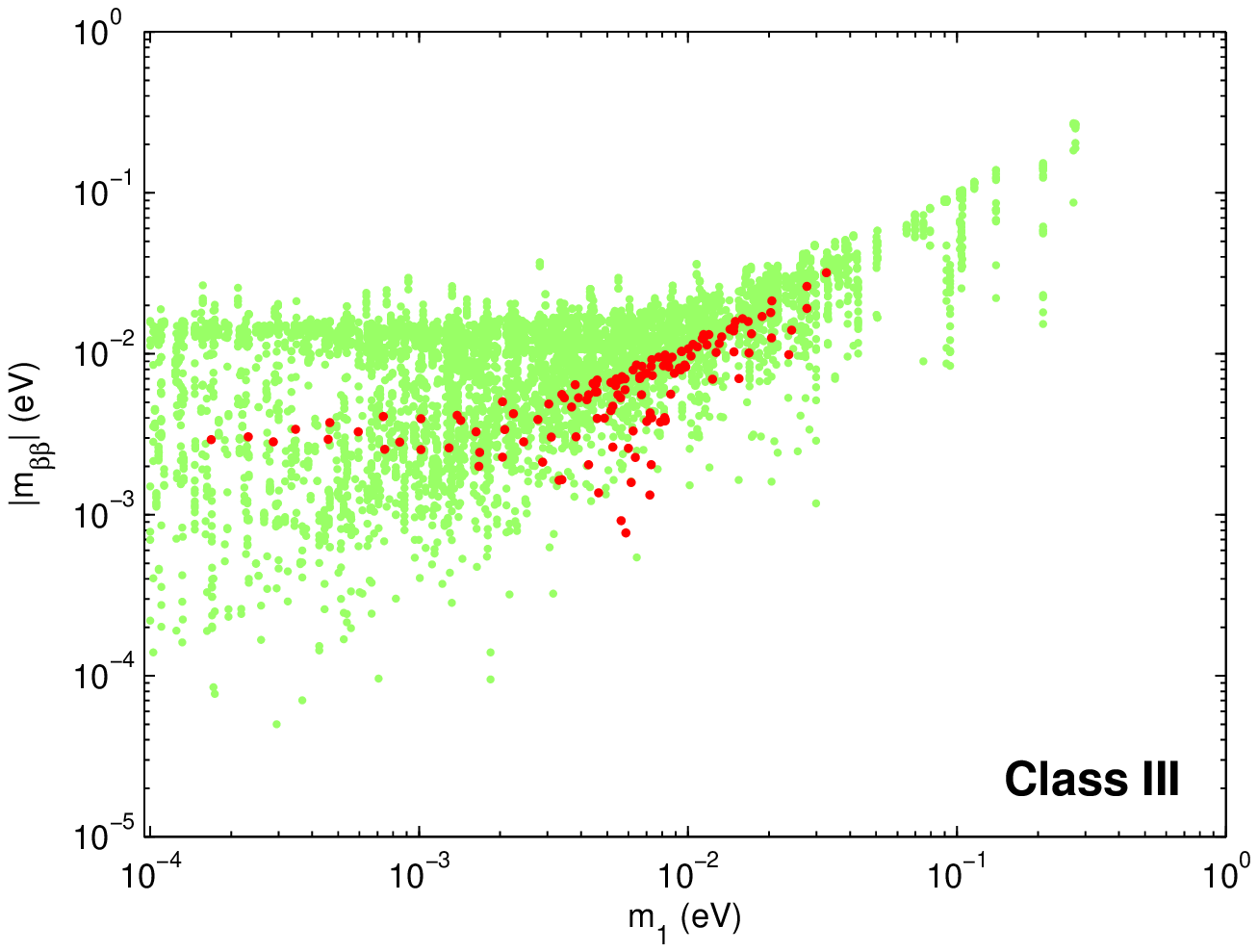}\\
\end{tabular}
\caption{Class III (normal hierarchy): The atmospheric mixing parameter $\sin^2\theta_{23}$ as a function of $|U_{e3}|$ and the effective Majorana mass $|m_{\beta\beta}|$ in terms of the lightest neutrino mass $m_1$. The darker (red) points correspond to the points which verify the experimental constraints.}
\label{fig:c3n}
\end{figure}

\begin{figure}[t]
\centering
\begin{tabular}{cc}
\includegraphics[width=0.5\linewidth]{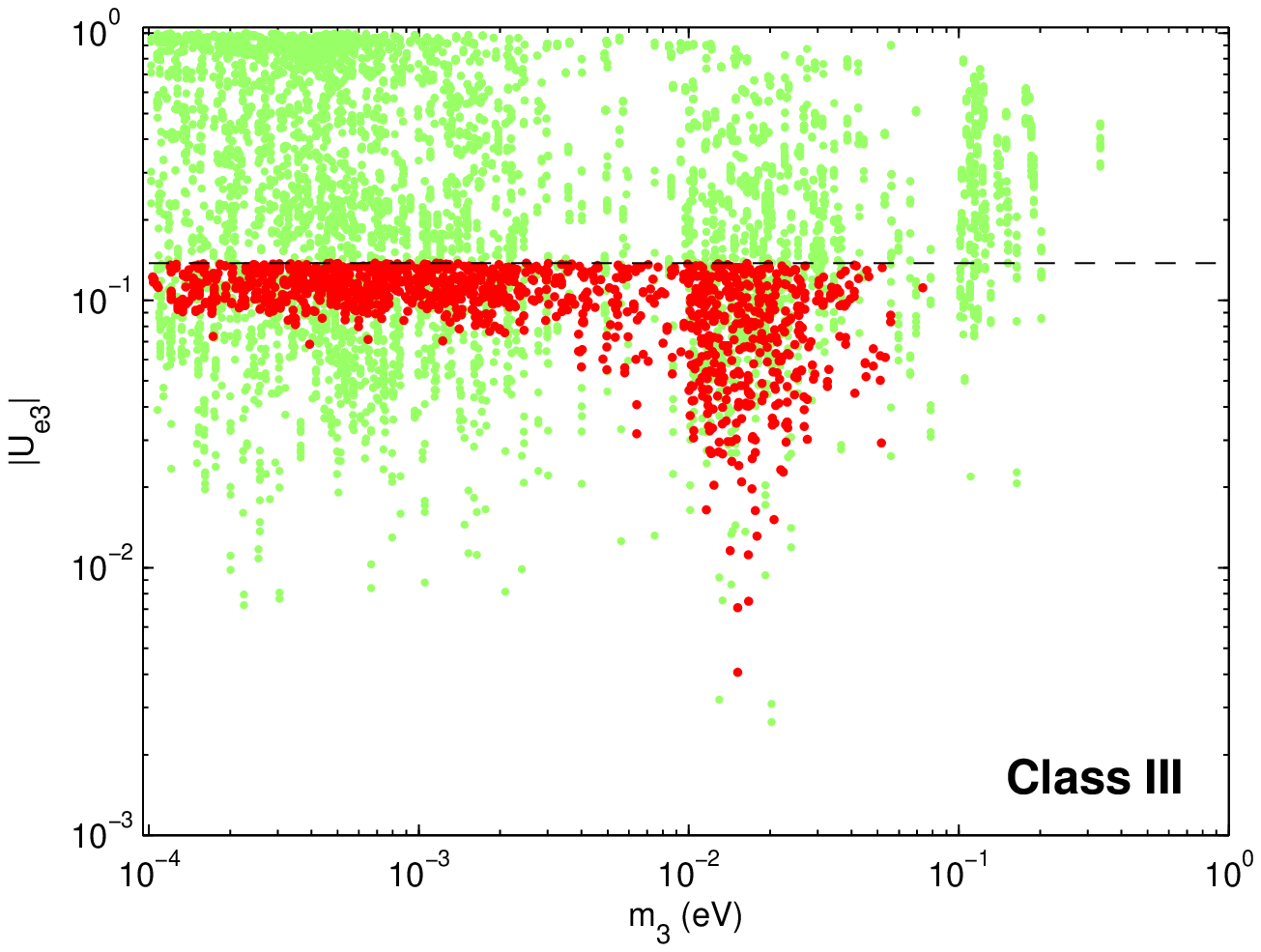} &
\includegraphics[width=0.5\linewidth]{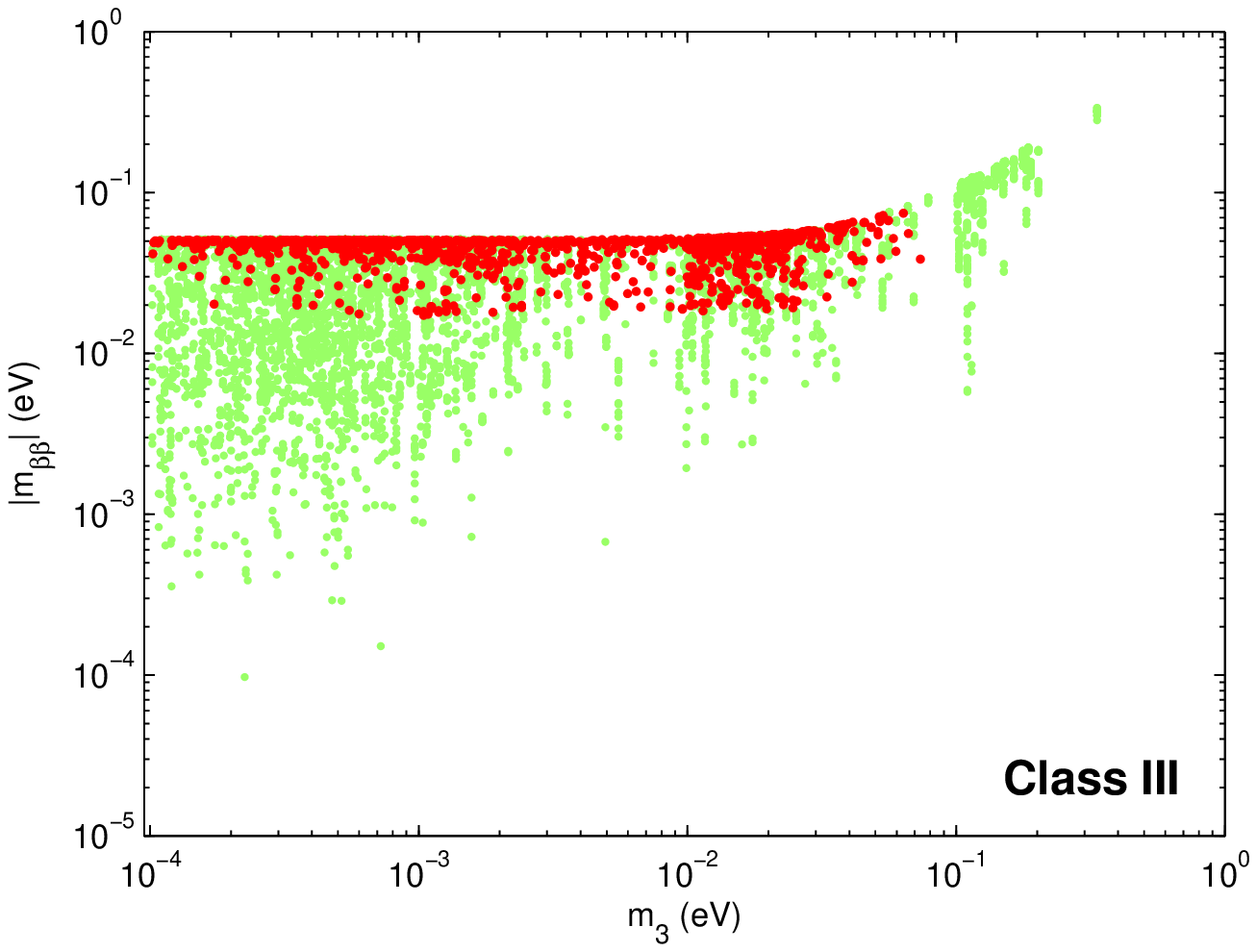}\\
\end{tabular}
\caption{Class III (inverted hierarchy): The mixing parameter $|U_{e3}|$ and the effective Majorana mass $|m_{\beta\beta}|$ in terms of the lightest neutrino mass $m_1$. The darker points define the parameter space consistent with the experimental constraints.}
\label{fig:c3i}
\end{figure}

The numerical results corresponding to the full complex case, including all physical phases, are presented in Fig.~\ref{fig:c3n} for a normal hierarchy of the neutrino masses and in Fig.~\ref{fig:c3i} for an inverted hierarchy. The allowed range for the lightest neutrino mass $m_1$ in the case of normal hierarchy is $10^{-4}\,\text{eV} \lesssim m_1\lesssim 3\times10^{-2}\,\text{eV}$. For the inverted hierarchy, only an upper bound can be established, $m_3\lesssim 7\times10^{-2}\,\text{eV}$.

As in the case of class II Ans\"atze, we can see that both scenarios are experimentally allowed. Again, the quantity $m_{\beta\beta}$ does not necessarily vanish. In the case of normal hierarchy we obtain the bounds $8\times 10^{-4}\,\text{eV} \lesssim |m_{\beta\beta}| \lesssim 3\times 10^{-2}\,\text{eV}$, which could in principle be tested in neutrinoless double beta decay experiments~\cite{KlapdorKleingrothaus:2004wj}. In the case of inverted hierarchy, we find the bounds $2\times 10^{-2}\,\text{eV} \lesssim |m_{\beta\beta}| \lesssim 7\times 10^{-2}\,\text{eV}$,  a range of values which is also at the reach of future neutrinoless double beta decays experiments \cite{Ejiri:1999rk}. We also obtain bounds on the mixing parameter $|U_{e3}|$, namely, $|U_{e3}| \gtrsim 9\times10^{-3}$ for normal hierarchy and $|U_{e3}| \gtrsim 4\times10^{-3}$ for the inverted hierarchy case.

From the above results we conclude that the leptonic mass matrices which belong to different classes have distinct features. In Table~\ref{tab:summary} we summarise some of the predictions of Class I, II and III Ans\"{a}tze.

\begin{table}[t]
 \centering
\begin{tabular}{ccc}
\textbf{Class} & \textbf{Normal Hierarchy} & \textbf{Inverted Hierarchy}
\\\hline
I   &
$\begin{array}{c}
m_1\gtrsim 8\times10^{-4}\,\text{eV}\\
|m_{\beta\beta}|\gtrsim 7\times10^{-5}\,\text{eV}\,\\
|U_{e3}|\gtrsim 4\times10^{-4}\\
\end{array}
$ & Excluded (unless $d_\nu \simeq m_3 \gtrsim 0.04$)\\\hline
II &
$\begin{array}{c}
m_1 \gtrsim 0.01\,\text{eV} \\
|m_{\beta\beta}| \gtrsim  0.01\,\text{eV}\\
|U_{e3}| \gtrsim 7\times10^{-3}\\
\end{array}
$ &
$\begin{array}{c}
m_3 \,\,\, \text{unrestricted}\\
|m_{\beta\beta}|\gtrsim 0.02\,\text{eV}\\
|U_{e3}| \gtrsim 0.01\\
\end{array}
$ \\\hline
III  &
$\begin{array}{c}
10^{-4}\,\text{eV}\lesssim m_1 \lesssim 0.03\,\text{eV}\\
8\times10^{-4}\,\text{eV} \lesssim |m_{\beta\beta}| \lesssim 0.03\,\text{eV}\\
|U_{e3}| \gtrsim 9\times10^{-3}\\
\end{array}
$ & $\begin{array}{c}
m_3\lesssim 0.07\,\text{eV}\\
0.02\,\text{eV}\lesssim|m_{\beta\beta}|\lesssim 0.07\,\text{eV} \\
|U_{e3}| \gtrsim 4\times10^{-3}\\
\end{array}
$\\\hline\\[1mm]
\end{tabular}
\caption{Summary of Ans\"{a}tze predictions: class I, II and III in the full complex case. A normal and an inverted hierarchy for the neutrino mass spectrum have been considered.}
\label{tab:summary}
\end{table}

\section{Summary and Conclusions}

We have emphasised that in any search for the experimentally viable texture-zero structures for fermion mass matrices, it is crucial to take into account the freedom to make WB transformations which do not change the physical content of a given structure but alter its form. In the case of the leptonic sector with Majorana neutrinos, we have investigated what zeros can be obtained, starting from arbitrary mass matrices for charged leptons and neutrinos, using the freedom to make WB transformations. In particular, we have shown that without loss of generality, one can choose a WB where $m_{\ell}$ is Hermitian and $m_{\ell}$, $m_{\nu}$ contain a total of three independent texture-zeros.

We have then classified and analysed the four texture-zero Ans\"atze for $m_{\ell}$ and $m_{\nu}$ with a parallel structure. These Ans\"atze do have physical implications, since not all the zeros can be obtained simultaneously, just by making WB transformations. It was shown that these four texture-zeros Ans\"atze can be classified in four classes, one of which is not compatible with the experimental data. The main predictions of these viable classes are summarised in Table~\ref{tab:summary}. We have also analysed how the predictions of these Ans\"atze differ from those studied by Frampton, Glashow and Marfatia, where the mass matrices $m_{\ell}$ and $m_{\nu}$ do not have a parallel structure.

\section*{Acknowledgements}
We thank Mariam Tortola for valuable discussions. G.C.B. would like to thank
Andrzej J. Buras for the kind hospitality at TUM (Munich, Germany). G.C.B. and
D.E.C are grateful for the warm hospitality of the CERN Physics Department (PH)
Theory (TH). This work was partially supported by Funda\c c\~ ao para a Ci\^
encia e a  Tecnologia (FCT, Portugal) through the projects POCTI/FNU/
44409/2002, PDCT/FP/63914/2005, PDCT/FP/63912/2005 and CFTP-FCT UNIT 777,
partially funded by POCTI (FEDER). The work of D.E.C. is presently supported by
a CFTP-FCT UNIT 777 fellowship. The work of G.C.B. was also supported by the
Alexander von Humboldt Foundation through a Humboldt Research Award.

\end{document}